\documentclass[12pt,a4paper]{article}
\usepackage{amssymb,amscd, amsmath}
\usepackage{amsthm}
\newtheorem{lemma}{Lemma}

\newtheorem{thm}{Theorem}
\def\Tr{{\mathrm{Tr}}}
\def\Prob{{\mathrm{Prob}}}
\def\<{\langle}
\def\>{\rangle}

\def\supp{\mathrm{supp}\,}
\def\BS{{\rm BS}}

\def\bbbc{\mathbb{C}}
\def\eps{\varepsilon}

\def\iH{{\cal H}}
\def\iK{{\cal K}}
\def\iM{{\cal M}}

\def\iX{{\cal X}}
\def\ot{\otimes}
\def\proof{{\it Proof:}\,\,}
\def\bN{\mathbb{N}}
\def\bbbn{\mathbb{N}}
\def\bE{\mathbb{E}}

\def\bfx{\mathbf{x}}
\def\bfy{\mathbf{y}}

\def\1{\mathbf{1}}
\def\Diag{\mathrm{Diag}}

\begin{document}
\ \vskip 1cm \centerline{\LARGE {\bf A limit relation for  entropy}} \bigskip
\centerline{\LARGE {\bf and channel capacity per unit cost}}
\bigskip
\bigskip
\bigskip
\centerline{\large Imre Csisz\'ar\footnote{E-mail: csiszar@renyi.hu.
Partially supported by the Hungarian Research Grant OTKA T068258.}$^{,4}$,
Fumio Hiai\footnote{E-mail: hiai@math.is.tohoku.ac.jp. Partially supported 
by Grant-in-Aid for Scientific Research (B)17340043.}$^{,5}$ and
D\'enes Petz\footnote{E-mail: petz@math.bme.hu.
Partially supported by the Hungarian Research Grant OTKA T068258.}$^{,4}$
}
\bigskip
\begin{center}
$^4$ Alfr\'ed R\'enyi Institute of Mathematics, \\H-1364 Budapest,
POB 127, Hungary
\end{center}
\begin{center}
$^5$ Graduate School of Information Sciences, Tohoku University \\
Aoba-ku, Sendai 980-8579, Japan
\end{center}
\bigskip
\bigskip\bigskip

\begin{quote}
{\bf Abstract:} In a quantum mechanical model, Di\'osi, Feldmann and Kosloff arrived
at a conjecture stating that the limit of the entropy of certain mixtures is the
relative entropy as system size goes to infinity. The conjecture is proven in this
paper for density matrices. The first proof is analytic and uses the quantum law of
large numbers. The second one clarifies the relation to channel capacity
per  unit cost for classical-quantum channels. Both proofs lead to generalizations 
of the conjecture.

\end{quote}
\bigskip
\bigskip
\section{Introduction}
It was conjectured by Di\'osi, Feldmann and Kosloff in \cite{diosi},
based on thermodynamical considerations, that the von Neumann entropy
of a quantum state equal to a mixture
$$
R_n:=\frac{1}{n}\left(\sigma\ot \rho^{\ot (n-1)}
+\rho \ot \sigma \ot \rho^{\ot (n-2)}+  \dots
+ \rho^{\ot (n-1)} \ot \sigma\right) 
$$
exceeds the entropy of a component asymptotically by the Umegaki relative
entropy $S(\sigma\|\rho)$, that is,
\begin{equation}\label{conj}
S(R_n)- (n-1)S(\rho)-S(\sigma)\to S(\sigma\|\rho)
\end{equation} as $n \to \infty$. Here $\rho$ and $\sigma$ are density 
matrices acting on a finite dimensional Hilbert space. Recall that
$S(\sigma)=-\Tr\, \sigma \log \sigma$ and
$$
S(\sigma \| \rho)=\left\{ \begin{array}{ll}
\Tr\, \sigma(\log \sigma -\log \rho) \quad & \hbox{if } \supp
\sigma\,\,\le \,\supp \rho\\
+\infty &\hbox{otherwise.} \end{array} \right. $$
Concerning the background of quantum entropy quantities, we refer to \cite{OP, pd}.

Apparently no exact proof of (\ref{conj}) has been published even for the classical
case, although for that case a heuristic proof is offered in \cite{diosi}.

In the paper first an analytic proof of (\ref{conj}) is given for the case
$\supp \sigma\,\,\le \,\supp \rho$, using an inequality between the
Umegaki and the Belavkin-Staszewski relative entropies, and the weak law of large
numbers in the quantum case. In the second part of the paper, it is clarified that
the problem is related to the theory of classical-quantum channels. The essential
observation is the fact that $S(R_n)- (n-1)S(\rho)-S(\sigma)$ in the conjecture is
a Holevo quantity (classical-quantum mutual information) for a certain channel for
which the relative entropy emerges as the  capacity per unit cost.

The two different proofs lead to two different generalizations of the conjecture.

\section{An analytic proof of the conjecture}

In this section we assume that $\supp \sigma \le  \supp \rho$ for the support
projections of $\sigma$ and $\rho$. One can simply compute:
\begin{align*}
S(R_n\|\rho^{\otimes n})
&=\Tr (R_n\log R_n-R_n\log\rho^{\otimes n}) \\
&=-S(R_n)-(n-1)\Tr\,\rho\log\rho-\Tr\,\sigma\log\rho.
\end{align*}
Hence the identity 
$$
S(R_n\|\rho^{\otimes n}) =-S(R_n)+(n-1)S(\rho)+S(\sigma\|\rho)+S(\sigma)
$$
holds. It follows that the conjecture (\ref{conj}) is equivalent to the
statement
$$
S(R_n\|\rho^{\otimes n}) \to 0 \quad\mbox{as}\quad n\to\infty
$$
when $\supp \sigma \le \supp \rho$.

Recall the Belavkin-Staszewski relative entropy
$$
S_\BS (\omega\|\rho)= \Tr (\omega \log (\omega^{1/2} \rho^{-1} \omega^{1/2}))
= - \Tr (\rho\, \eta (\rho^{-1/2}\omega \rho^{-1/2}))
$$
if $\supp \omega \le \supp \rho$, where $\eta(t):=-t \log t$, see 
\cite{BS, OP}. It was proved by Hiai and Petz that
\begin{equation}\label{HPin}
S (\omega\|\rho) \le S_\BS (\omega\|\rho),
\end{equation}
see \cite{HP}, or Proposition 7.11 in \cite{OP}.

\begin{thm}\label{T:1}
If $\supp \sigma \le \supp \rho$, then $S(R_n) -(n-1)S(\rho)-S(\sigma)\to 
S(\sigma \| \rho)$ as $n \to \infty$.
\end{thm}

\proof
We want to use the quantum law of large numbers, see Proposition 1.17
in \cite{OP}. Assume that $\rho$ and $\sigma$ are $d \times d$ density matrices
and we may suppose that $\rho$ is invertible. Due to the GNS-construction
with respect to the limit $\varphi_\infty$ of the product states $\varphi_n(A)
=\Tr\,\rho^{\ot n}A$ on the $n$-fold tensor product $M_d(\bbbc)^{\ot n}$,
$n \in \bbbn$, all finite tensor products $M_d(\bbbc)^{\ot n}$ are embedded into
a von Neumann algebra $\iM$ acting on a Hilbert space $\iH$. If $\gamma$ denotes
the right shift and $X:=\rho^{-1/2}\sigma\rho^{-1/2}$, then $R_n$ is written as
$$
R_n=(\rho^{1/2})^{\otimes n}\Biggl({1\over n}\sum_{i=0}^{n-1}\gamma^i(X)\Biggr)
(\rho^{1/2})^{\otimes n}.
$$

By  inequality (\ref{HPin}), we get
\begin{align}
0\le S(R_n\|\rho^{\otimes n})
&\le S_{\mathrm{BS}}(R_n\|\rho^{\otimes n}) \nonumber \\
&=-\Tr \Bigl(\rho^{\otimes n}\,
\eta\Bigl((\rho^{-1/2})^{\otimes n}R_n(\rho^{-1/2})^{\otimes n}\Bigr)\Bigr)
\nonumber\\
&=\Big\< \Omega ,\eta\Biggl({1\over n}\sum_{i=0}^{n-1}
\gamma^i(X)\Biggr)\Omega\Big\>\,,\label{upb}
\end{align}
where $\Omega$ is the cyclic vector in the GNS-construction.

The law of large numbers gives
$$
{1\over n}\sum_{i=0}^{n-1}\gamma^i(X) \to I $$
in the strong operator topology in $B(\iH)$, since $\varphi(X)=\Tr\,\rho\rho^{-1/2}
\sigma\rho^{-1/2}=1$.

Since the continuous functional calculus preserves the strong convergence
(simply due to approximation by polynomials on a compact set), we obtain
$$
\eta\Biggl({1\over n}\sum_{i=0}^{n-1}\gamma^i(X)\Biggr)
\to \eta(I)=0\ \ \mbox{strongly}.
$$
This shows that the upper bound (\ref{upb}) converges to 0 and the
proof is complete. \qed

By the same proof one can obtain that for
$$
R_{m,n}:=\frac{1}{n}\left(\sigma^{\otimes m}\ot \rho^{\ot(n-1)}
+\rho \ot \sigma^{\otimes m} \ot \rho^{\ot(n-2)}
+  \dots  + \rho^{\ot(n-1)} \ot \sigma^{\otimes m}\right),
$$
the limit relation
\begin{equation}
S(R_{m,n})-(n-1)S(\rho)-mS(\sigma) \to m S(\sigma\|\rho)
\end{equation}
holds as $n\to\infty$ when $m$ is fixed.

In the next theorem we treat the probabilistic case in a matrix language.
The proof includes the case when  $\supp \sigma \le \supp \rho$ is not true. Those
readers who are not familiar with the quantum setting of the previous theorem are
suggested to follow the arguments below.

\begin{thm} \label{T:2}
Assume that $\rho$ and $\sigma$ are commuting density matrices. Then
$S(R_n) -(n-1)S(\rho)-S(\sigma)\to S(\sigma \| \rho)$ as $n \to \infty$.
\end{thm}

\proof
We may assume that $\rho=\Diag(\mu_1,\dots,\mu_\ell, 0,\dots,0)$ and
$\sigma=\Diag(\lambda_1,\dots,\lambda_d)$ are $d \times d$ diagonal matrices,
$\mu_1, \dots, \mu_\ell >0$ and $\ell < d$. (We may consider
$\rho,\sigma$ in a matrix algebra of bigger size if $\rho$ is invertible.) 
If $\supp \sigma \le \supp \rho$, then $\lambda_{\ell+1}= \dots =\lambda_d=0$;
this will be called the regular case. When $\supp \sigma \le \supp \rho$ is not true,
we may assume that $\lambda_d>0$ and we refer to the singular case.

The eigenvalues of $R_n$ correspond to elements $(i_1,\dots ,i_n)$ of
$\{1,\dots,d\}^n$:
\begin{equation}\label{ev}
{1\over n}(\lambda_{i_1}\mu_{i_2}\cdots\mu_{i_n}
+\mu_{i_1}\lambda_{i_2}\mu_{i_3}\cdots\mu_{i_n}
+\dots+\mu_{i_1}\cdots\mu_{i_{n-1}}\lambda_{i_n}). \end{equation}
We divide the eigenvalues in three different groups as follows:
\begin{itemize}
\item[(a)] $A$ corresponds to $(i_1,\dots ,i_n)\in\{1,\dots,d\}^n$ with
$1 \le i_1,\dots ,i_n \le \ell$,
\item[(b)] $B$ corresponds to $(i_1,\dots ,i_n)\in\{1,\dots,d\}^n$ which contains
exactly one $d$,
\item[(c)] $C$ is the rest of the eigenvalues. \end{itemize}

If the eigenvalue (\ref{ev}) is in group $A$, then it is
$$
{(\lambda_{i_1}/\mu_{i_1})+\dots+(\lambda_{i_n}/\mu_{i_n})\over n}
\,\mu_{i_1}\mu_{i_2}\cdots\mu_{i_n}.
$$
First we compute
$$
\sum_{\kappa \in A}\eta(\kappa)=\sum_{i_1,\dots,i_n} \eta
\biggl({(\lambda_{i_1}/\mu_{i_1})+\dots+(\lambda_{i_n}/\mu_{i_n})\over n}
\,\mu_{i_1}\cdots\mu_{i_n}\biggr).
$$
Below the summations are over $1\le i_1,\dots,i_n \le \ell$:
\begin{align*} 
&\sum_{i_1,\dots,i_n} \eta
\biggl({(\lambda_{i_1}/\mu_{i_1})+\dots+(\lambda_{i_n}/\mu_{i_n})\over n}
\,\mu_{i_1}\cdots\mu_{i_n}\biggr) \\ & \qquad =-\sum_{i_1,\dots,i_n}
({(\lambda_{i_1}/\mu_{i_1})+\dots+(\lambda_{i_n}/\mu_{i_n})\over n}
\,\mu_{i_1}\cdots\mu_{i_n}\biggr)\log(\mu_{i_1}\cdots\mu_{i_n})+Q_n \\
&\qquad =-{1\over n}\sum_{k=1}^n\Biggl(\sum_{i_1,\dots,i_n}
\lambda_{i_1}\mu_{i_2}\cdots\mu_{i_n}\log\mu_{i_k}
+\sum_{i_1,\dots,i_n}\lambda_{i_1}\mu_{i_2}\cdots\mu_{i_n}\log\mu_{i_k} \\
&\qquad\qquad\qquad\qquad\qquad+\dots
+\sum_{i_1,\dots,i_n}\lambda_{i_1}\mu_{i_2}\cdots\mu_{i_n}\log\mu_{i_k}
\Biggr)+Q_n \\
&\qquad =-{1\over n}\sum_{k=1}^n\Biggl((n-1)\sum_{i_k}\mu_{i_k}\log\mu_{i_k}
+\sum_{i_k}\lambda_{i_k}\log\mu_{i_k}\Biggr)+Q_n \\
&\qquad =(n-1)S(\rho)-\sum_{i=1}^\ell\lambda_i\log\mu_i +Q_n,
\end{align*}
where
$$
Q_n:=\sum_{i_1,\dots,i_n}(\mu_{i_1}\cdots\mu_{i_n})
\eta\biggl({(\lambda_{i_1}/\mu_{i_1})+\dots+(\lambda_{i_n}/\mu_{i_n})\over n}
\biggr).
$$

Consider a probability space
$$
(\Omega,\mathbb{P}):=\bigl(\{1,\dots,\ell\}^\bN,(\mu_1,\dots,\mu_\ell)^\bN\bigr),
$$
where $(\mu_1,\dots,\mu_\ell)^\bN$ is the product of the measure on $\{1,
\dots,\ell\}$ with the distribution $(\mu_1,\dots,\mu_\ell)$. For each $n\in\bN$
let $X_n$ be a random variable on $\Omega$ depending on the $n$th $\{1,\dots,\ell\}$
so that the value of $X_n$ at $i\in\{1,\dots,\ell\}$ is $\lambda_i/\mu_i$. Then
$X_1,X_2,\dots$ are identically distributed independent random variables and $Q_n$
is the expectation value
of
$$
\eta \biggl({X_1+\dots+X_n\over n}\biggr).
$$
The strong law of large numbers says that
$$
{X_1+\dots+X_n\over n}\to \bE(X_1)
=\sum_{i=1}^\ell\biggl({\lambda_i\over\mu_i}\biggr)\mu_i=\sum_{i=1}^\ell \lambda_i
\ \ \mbox{almost surely}.
$$
Since $ \eta((X_1+\dots+X_n)/n)$ is uniformly bounded, the Lebesgue bounded
convergence theorem implies that
$$
Q_n\to \eta\Big(\sum_{i=1}^\ell \lambda_i\Big) $$
as $n\to\infty$.

In the regular case $\sum_{i=1}^\ell \lambda_i=1$, $Q_n \to 0$ and all
non-zero eigenvalues are in group $A$. Hence we have
$$
S(R_n)-(n-1)S(\rho)-S(\sigma)
=-\sum_{i=1}^\ell\lambda_i\log\mu_i+ \sum_{i=1}^\ell \lambda_i\log\lambda_i +Q_n
=S(\sigma\|\rho)+Q_n
$$
and the statement is clear.

Next we consider the singular case, when we have
$$
\sum_{\kappa \in A} \eta(\kappa)=(n-1)S(\rho)+ O(1),
$$
and we turn to eigenvalues in $B$. If the eigenvalue corresponding to
$(i_1,\dots ,i_n)\in\{1,\dots,d\}^n$ is in group $B$ and $i_1=d$, then
the eigenvalue is
$$
{1\over n}\lambda_{d}\mu_{i_2}\dots\mu_{i_n}.
$$
It follows that
\begin{eqnarray*}
&& -\sum_{i_2,\dots,i_n}
\Big(\frac{\lambda_d\mu_{i_2}\cdots\mu_{i_n}}{n}\Big)
\log \Big(\frac{\lambda_d\mu_{i_2}\cdots\mu_{i_n}}{n}\Big)
\cr
&&\qquad = -\frac{\lambda_d}{n}\sum_{i_2,\dots,i_n}
(\mu_{i_2}\cdots\mu_{i_n}) \log (\mu_{i_2}\cdots\mu_{i_n})
-\frac{\lambda_d}{n}\log \frac{\lambda_d}{n}
\cr
&&\qquad
= \frac{\lambda_d}{n}(n-1)S(\rho)-\frac{\lambda_d}{n}\log \frac{\lambda_d}{n}.
\end{eqnarray*}
When $i_2=d, \dots, i_n=d$, we get the same quantity, so this should be
multiplied with $n$:
$$
\sum_{\kappa \in B} \eta(\kappa)=
\lambda_d (n-1)S(\rho)- \lambda_d \log \frac{\lambda_d}{n}.
$$
We make a lower estimate to the entropy of $R_n$ in such a way that we
compute $\sum_\kappa \eta(\kappa)$ when $\kappa$ runs over $A$ and $B$.
It is clear now that
\begin{eqnarray*}
S(R_n)- (n-1)S(\rho)-S(\sigma)
&\ge& \sum_{\kappa \in A} \eta(\kappa)+\sum_{\kappa \in B} \eta(\kappa)-
(n-1)S(\rho)-S(\sigma) \cr
&\ge& \lambda_d (n-1)S(\rho)+\lambda_d \log n + O(1)
\to +\infty
\end{eqnarray*}
as $n \to \infty$.\qed

\section{Interpretation as capacity}

A classical-quantum channel with classical input alphabet $\iX$ transfers the input
$x\in \iX$ into the output $W(x) \equiv \rho_x$ which is a density matrix acting
on a Hilbert space $\iK$. We restrict ourselves to the case when $\iX$ is finite and
$\iK$ is finite dimensional.

If a classical random variable $X$ is chosen to be the input, with probability
distribution $P=\{p(x): x \in \iX\}$, then the corresponding output is the
quantum state $\rho_X:= \sum_{x \in \iX} p(x) \rho_x$. When a measurement
is performed on the output quantum system, it gives rise to an output
random variable $Y$ which is jointly distributed with the input $X$. If a partition
of unity $\{F_y :y \in \iX\}$ in $B(\iK)$ describes the measurement, then
\begin{equation}\label{E:tr}
\Prob (Y=y\, |\, X=x )= \Tr\, \rho_x F_y \qquad (x,y \in \iX).
\end{equation}

According to the Holevo bound, we have
\begin{equation}\label{E:Ho}
I(X\wedge Y):=H(Y)-H(Y|X) \le I(X,W):=S(\rho_X)-\sum_{x \in \iX} p(x) S(\rho_x),
\end{equation}
which is actually a simple consequence of the monotonicity of the relative entropy
under state transformation \cite{Ho1973b}, see also \cite{OPW}. $I(X,W)$ is the
so-called Holevo quantity or classical-quantum  mutual information, and it satisfies
the identity
\begin{equation}\label{E:ident}
\sum_{x \in \iX} p(x) S(\rho_x\|\rho)=I(X,W)+S(\rho_X\|\rho),
\end{equation}
where $\rho$ is an arbitrary density.

The channel is used to transfer sequences from the classical alphabet;
$\bfx=(x_1,x_2,\dots,x_n) \in \iX^n$ is transferred into the quantum state
$W^{\ot n}(\bfx)=\rho_\bfx:=\rho_{x_1}\ot \rho_{x_2}\ot \dots \ot \rho_{x_n}$.
A code for the channel $W^{\ot n}$ is defined by a subset $A_n \subset \iX^n$,
which is called a codeword set. The decoder is a measurement
$\{F_\bfy: \bfy \in \iX^n\}$. The probability of error is $\Prob(X\ne Y)$, where
$X$ is the input random variable uniformly distributed on $A_n$ and the output
random variable is determined by (\ref{E:tr}), where $x$ and $y$ are replaced by
$\bfx$ and $\bfy$.

The essential observation is the fact that $S(R_n)- (n-1)S(\rho)-S(\sigma)$
in the conjecture is a Holevo quantity in case of a channel with input
sequences $(x_1,x_2,\dots, x_n)\in \{0,1\}^n$ and outputs
$\rho_{x_1}\ot \rho_{x_2}\ot \dots \ot\rho_{x_n}$, where $\rho_0=\sigma$,
$\rho_1=\rho$ and the codewords are all sequences containing exactly one 0. More
generally, we shall consider Holevo quantities
$$
I(A,\rho_0,\rho_1):= S\Big(\frac{1}{|A|}\sum_{\bfx \in A}
\rho_\bfx\Big)-\frac{1}{|A|}\sum_{\bfx \in A}
S(\rho_\bfx).
$$
defined for any set $A \subset \{0,1\}^n$ of binary sequences of length $n$.

The concept related to the conjecture we study is the channel capacity per
unit cost which is defined next for simplicity only in the case where
$\iX=\{0,1\}$, the cost of a character $0\in \iX$ is 1, while the cost of
$1 \in \iX$ is 0.

For a memoryless channel with a binary input alphabet $\iX=\{0,1\}$ and
an $\eps>0$, a number $R>0$ is called an $\eps$-achievable
rate per unit cost if for every $\delta >0$ and for any sufficiently large $T$, 
there exists a code of length $n>T$ with at least $e^{T(R-\delta)}$
codewords such that each of the codewords contains at most $T$ 0's and the error
probability is at most $\eps$. The largest $R$ which is an $\eps$-achievable
per unit cost for every $\eps>0$ is the channel capacity per unit cost.

\begin{lemma}\label{L:2}
For an arbitrary $A \subset \{0,1\}^n$,
$$
I(A,\rho_0,\rho_1) \le c(A) S(\rho_0\|\rho_1)
$$
holds, where
$$
c(A):= \frac{1}{|A|}\sum_{\bfx \in A}|\{i: x_i=0\}|.
$$
\end{lemma}

\proof
Let $c(\bfx):=|\{i: x_i=0\}|$ for $\bfx \in A$. Since 
$I(A,\rho_0,\rho_1)$ is a particular Holevo quantity $I(X,  W^{\ot n})$,
we can use the identity (\ref{E:ident}) to get an upper bound
$$
\frac{1}{|A|}\sum_{\bfx \in A} S(\rho_\bfx\|\rho_1^{\ot n})=
\frac{1}{|A|}\sum_{\bfx \in A} c(\bfx)S(\rho_0\|\rho_1)=
c(A)S(\rho_0\|\rho_1)
$$
for $I(A,\rho_0,\rho_1)$. \qed

\begin{lemma}\label{L:4}
 If $A \subset \{0,1\}^n$ is a code of the channel $W^{\ot n}$,
whose probability of error $($for some decoding scheme$)$ does not exceed a given
$0 < \eps <1$, then
$$
(1-\eps) \log |A| - \log 2 \le I(A,\rho_0,\rho_1).  
$$
\end{lemma}
\proof
The right-hand side is   a bound for the classical mutual information
$I(X\wedge Y)=H(Y)-H(Y|X)$, where $Y$ is the channel output, see (\ref{E:Ho}).
Since the error probability $\Prob(X\ne Y)$ is smaller than $\eps$, application
of the Fano inequality (see \cite{CT}) gives
$$
H(X |Y )\le \eps \log |A| +\log 2.
$$
Therefore
$$
I(X \wedge Y )= H(X)-H(X | Y ) \ge (1-\eps)\log |A| - \log2,
$$
and the proof is complete. \qed

The above two lemmas shows that the relative entropy $S(\rho_0\|\rho_1)$ is an
upper bound for the channel capacity per unit cost of the channel $W(0)=\rho_0$
and $W(1)=\rho_1$ with a binary input alphabet. In fact, assume that $R>0$ is an
$\eps$-achievable rate. For every $\delta>0$ and $T>0$ there is a code
$A\subset\{0,1\}^n$ for which we get by Lemmas \ref{L:2} and \ref{L:4}
\begin{align*}
TS(\rho_0\|\rho_1)&\ge c(A)S(\rho_0\|\rho_1)\ge I(A,\rho_0,\rho_1) \\
&\ge(1-\eps)\log|A|-\log2 \\
&\ge(1-\eps)T(R-\delta)-\log 2.
\end{align*}
Since $T$ is arbitrarily large and $\eps,\delta$ are arbitrarily small,
$R\le S(\rho_0\|\rho_1)$ follows. That $S(\rho_0\|\rho_1)$ equals the channel
capacity per unit cost will be verified below.

\begin{thm}\label{T:3}
Let the classical-quantum channel $W:\iX=\{0,1\} \to B(\iK)$ be defined as
$W(0)=\rho_0\equiv \sigma$ and $W(1)=\rho_1\equiv\rho$. Assume that
$A_n \subset \{0,1\}^n$ is chosen such that
\begin{itemize}
\item[(a)]
each element $\bfx=(x_1,x_2,\dots, x_n) \in A_n$ contains at most $\ell$
copies of 0,
\item[(b)]
$\log |A_n|/ \log n \to c$ as $n \to \infty$,
\item[(c)]
$$
c(A_n):= \frac{1}{|A_n|}\sum_{\bfx \in A_n}|\{i: x_i=0\}| \to c
\quad\mbox{as}\quad n \to \infty
$$
\end{itemize}
for some real number $c>0$ and for some natural number $\ell$. If the random variable
$X_n$ has a uniform distribution on $A_n$, then
$$
\lim_{n \to \infty} \Big( S(\rho_{X_n})-\frac{1}{|A_n|} \sum_{{\bf x} \in A_n}
S(\rho_{{\bf x}}) \Big) = cS(\sigma\|\rho).
$$
\end{thm}

The proof of the theorem is divided into lemmas. We need the direct
part of the so-called quantum Stein lemma obtained in \cite{HP},
see also \cite{Bje, Hay, ON, pd}.

\begin{lemma}\label{L:1}
Let $\rho_0$ and $\rho_1$ be density matrices. For every $\eta >0$ and
$0 < R < S(\rho_0\|\rho_1)$, if $N$ is sufficiently large, then there is
a projection $E \in B(\iK^{\ot N})$ such that
$$
\alpha_N[E]:=\Tr\, \rho_0^{\ot N}(I-E) < \eta
$$
and for $\beta_N[E]:=\Tr\, \rho_1^{\ot N}E$ the estimate
$$
\frac{1}{N} \log \beta_N[E] <  - R
$$
holds.
\end{lemma}

Note that $\alpha_N$ is called the error of the first kind, while $\beta_N$ is
the error of the second kind.

\begin{lemma}\label{L:3}
Assume that $\eps >0$, $0<R< S(\rho_0 \|\rho_1)$, $\ell$ is a positive
integer and the sequences $\bfx$ in $A_n \subset \{0,1\}^n$ contain at most $\ell$
copies of 0. Let the codewords be the  $N$-fold repetitions $\bfx^N=(\bfx,\bfx,
\dots,\bfx)$  of the sequences $\bfx \in A_n$. If $N$ is the integer part of
$$
\frac{1}{R} \log \frac{2n}{\eps}
$$
and $n$ is large enough, then there is a decoding scheme such that the error 
probability is smaller than $\eps$.

\end{lemma}

\proof
We follow the probabilistic construction in \cite{Ve}.
Let the codewords be the $N$-fold repetitions $\bfx^N=(\bfx,\bfx,\dots,\bfx)$ of
the sequences $\bfx \in A_n$. The corresponding output density matrices
act on the Hilbert space $\iK^{\ot Nn}\equiv (\iK^{\ot n})^{\ot N}$. We
decompose this Hilbert space into an $N$-fold product in a different way.
For each $1 \le i \le n$, let $\iK_i$ be the tensor product of the factors
$i,i+n,i+2n,\dots, i+(N-1)n$. So $\iK$ is identified with $\iK_1 \ot \iK_2
\ot \dots \ot \iK_n$.

For each $1 \le i \le n$ we perform a hypothesis testing on the Hilbert space
$\iK_i$. The 0-hypothesis is that the $i$th component of the
actually chosen $\bfx \in A_n$ is $0$. Based on the channel outputs at
time instances $i,i+n,\dots, i+(N-1)n$, the  0-hypothesis is tested against the
alternative hypothesis that the $i$th component of $\bfx$
is $1$. According to the quantum Stein lemma (Lemma \ref{L:1}),
given any $\eta>0$ and $0<R < S(\sigma\|\rho)$, for $N$ sufficiently
large, there exists a test $E_i$ such that the probability of error of
the first kind is smaller than $\eta$, while the probability of error of
the second kind is smaller than $e^{-NR}$. The projections $E_i$ and $I-E_i$ form
a partition of unity in the Hilbert space $\iK_i$, and the $n$-fold tensor
product of these commuting projection will give a partition of unity
in $\iK^{\ot Nn}$. Let $\bfy \in \{0,1\}^n$ and set $F_{\bfy}:= \ot_{i=1}^n F_{y_i}$,
where $F_{y_i}=E_i$ if $y_i=0$ and $F_{y_i}=I-E_i$ if $y_i=1$. Therefore,
the result of decoding can be an arbitrary $0$--$1$ sequence in $\{0,1\}^n$.

The decoding scheme gives $\bfy \in \{0,1\}^n$ in such a way that $y_i=0$
if the tests accepted the 0-hypothesis for $i$ and $y_i=1$ if the alternative was
accepted. The error probability should be estimated:
\begin{align*}
\Prob(Y\ne X|X=\bfx)&=
\sum_{\bfy:\bfy\ne \bfx} \Tr\,\rho_\bfx^{\ot N} F_\bfy
=\sum_{\bfy:\bfy\ne \bfx}\prod_{i=1}^n \Tr\,\rho_{x_i}^{\ot N} F_{y_i}  \cr
&\le \sum_{i=1}^n \sum_{\bfy: y_i \ne x_i}
\prod_{j=1}^n \Tr\,\rho_{x_j}^{\ot N} F_{y_j}
\le \sum_{i=1}^n \Tr\,\rho_{x_i}^{\ot N} (I-F_{x_i}).
\end{align*}
If $x_i=0$, then
$$
\Tr\, \rho_{x_i}^{\ot N} (I-F_{x_i}) = \Tr\, \rho_{0}^{\ot N} (I-E_{i}) \le \eta,
$$
because it is an error of the first kind. When $x_i=1$,
$$
\Tr\, \rho_{x_i}^{\ot N} (I-F_{x_i}) = \Tr\, \rho_{1}^{\ot N} E_{i} \le e^{-R N}
$$
from the error of the second kind. It follows that
$\ell \eta +ne^{-NR}$ 
is a bound for the error probability. The first term will be small
if $\eta$ is small. The second term will be small if $N$ is large enough. If both
terms are majorized by $\eps/2$, then the statement of the lemma holds.
We can choose $n$ so large that $N$ defined by the statement should be
large enough. \qed

{\it Proof of Theorem \ref{T:3}}:
Since Lemma \ref{L:2} gives an upper bound, that is,
$$
\limsup_{n\to\infty}\Big( S(\rho_{X_n})-\frac{1}{|A_n|} \sum_{{\bf x} \in
A_n}
S(\rho_{{\bf x}}) \Big) \le cS(\sigma\|\rho),
$$
it remains to prove that
$$
\liminf_{n \to \infty} \Big( S(\rho_{X_n})-\frac{1}{|A_n|}
\sum_{{\bf x} \in A_n} S(\rho_{{\bf x}}) \Big) \ge cS(\sigma\|\rho).
$$

Lemma \ref{L:3} is about the $N$-times repeated input $X^N$ and describes
a decoding scheme with error probability at most $\eps$. According to
Lemma \ref{L:4} we have
$$
(1-\eps) \log |A_n| -1 \le S(\rho_{X^N})-\frac{1}{|A|}
\sum_{\bfx \in A_n} S(\rho_{\bfx^N}).
$$
 From the subadditivity of the entropy we have
$$
S(\rho_{X^N}) \le N S(\rho_{X})
$$
and
$$
S(\rho_{\bfx^N}) = N S(\rho_{\bfx})
$$
holds due to the additivity for product. It follows that
$$
(1-\eps) \frac{\log |A_n|}{N} -\frac{1}{N}
\le S(\rho_{X})-\frac{1}{|A_n|}\sum_{\bfx \in A_n} S(\rho_{\bfx}).
$$
 From the choice of $N$ in Lemma \ref{L:3} we have
$$
R \,\frac{\log |A_n|}{\log n}\,\frac{\log n}{\log n +\log 2 - \log \eps}
\le \frac{\log |A_n|}{N}
$$
and the lower bound is arbitrarily close to $cR$. Since
$R < S(\rho_0\| \rho_1)$ was arbitrary, the proof is complete. \qed

\end{document}